\documentclass{elsart}
\usepackage{amssymb,epsfig}
\journal{Physica E}
\begin{document}
\begin{frontmatter}
\title{Hybrid-impurity resonances in anisotropic quantum dots}
\author{V.A. Margulis}, \author{A.V. Shorokhov}
\address{Institute of Physics and Chemistry, Mordovian State
University, 430000 Saransk, Russia} \ead{alex.shorokhov@mail.ru}
\begin{abstract}
The absorption of electromagnetic radiation of an anisotropic
quantum dot is theoretically investigated taking into account the
processes associated with simultaneous scattering from ionized
impurities. It is shown that the scattering of electrons by
impurities leads to the resonance absorption even if we have only
one impurity in the quantum dot. Explicit formula is derived for
the absorption coefficient. The positions of the resonances peaks
are found. The effects of external magnetic field on the resonance
absorption are studied.
\end{abstract}
\end{frontmatter}

\section{Introduction}
In the past few years optical electron transitions in quantum dots
(QD) are widely studied \cite{Gru02} in connection to their
potential use in new or improved optoelectronic devices
\cite{Mic03}, such as quantum dots lasers or infrared
photodetectors \cite{Ryz04}. What is more, the investigation of
the intraband optical transitions offers an efficient method for
studying many fundamental physical phenomena \cite{Cal00}.

Quantum dots support several types of intraband resonances. In
particular, apart from direct absorption of electromagnetic
radiation, there can occur processes involving absorption
(emission) of a photon with simultaneous absorption (emission) of
a phonon or scattering from an impurity. The last processes, in
particular, can modify the selection rules in the optical
transitions and lead to losses in optical devices based on QDs. In
view of this, the study of carrier impurity interaction in QDs is
crucial for the development of technology of quantum dots
production. With the experimental realization of nanostructures
devices a lot of theoretical investigations of electron-impurity
interaction in nanostructures have been performed (see, for
example, \cite{Den98}-\cite{El94}). In particular, the effect of
repulsive scattering centers on the energy spectrum of a quantum
dot was studied in
\cite{Hal96,Zar02,Zhu94,Bet94,Cha01,Lim08,Bas07}.

Modern nanotechnology enables one to fabricate quantum dots of
different shapes. In the past years the significant interest has
been given to semiconductor quantum wells and quantum dots that
are characterized by an asymmetric confining potential
\cite{Ahn87,Yil06,Zha05}.

In this paper, we theoretically study the optical absorption of
anisotropic QDs subjected to a uniform magnetic field arbitrarily
directed with potential symmetry axis taking into account the
processes associated with simultaneous scattering from ionized
impurities. The applied magnetic field plays the important role in
the systems based on the QDs due to possibility to control both
the working frequency and the magnitude of intraband absorption
changing the amplitude and direction of the external magnetic
field.

We model the semiconductor QD with a asymmetric parabolic
confinement $V(\vec{r})=m^*(\Omega^{2}_x x^2 +
\Omega^{2}_yy^2+\Omega^{2}_z z^2)/2$ (here $m^*$ is the effective
mass of the electron, $\Omega_i$ ($i=x,y,z$) are the
characteristic frequencies of the parabolic potential). Note that
usually the confinement in the $z$ direction is much stronger that
in the $xy$ plane ($\Omega_x,\Omega_y\ll\Omega_z$).

In the case of resonances arising upon absorption of
electromagnetic radiation by electrons of a quantum dots with
simultaneous scattering from ionized impurities, the absorption
coefficient can be found by applying ordinary perturbation theory
for the interaction of electrons with the high frequency
electromagnetic field and the ionized impurity \cite{Mar04}. In
this case the absorption peaks are due to the selection rules for
the transitions in second-order perturbation theory and low of
conversation energy in such transitions. We assume that impurity
is located in the center of dot. Note that resonances of this type
(cyclotron-impurity resonances) in the bulk semiconductors have
been extensively studied both theoretically and experimentally
(see, e.g. \cite{Bas78,Dun84,Boh81}).

The screened potential of an ionized impurity located at the
coordinate origin can be represented by the standard relationship

\begin{equation}
\label{1} U(r)=\frac{ze^2}{\varepsilon r}e^{-kr}\;,
\end{equation}
Here $\varepsilon$ is the real part of the dielectric constant (we
suppose there is no dispersion in the frequency range considered
here), $ze$ is the impurity charge, $k=1/r_0$, where $r_0$ is the
screening radius. It is well-known that  $r_0$ is independent of
the magnetic field in the case of nondegenerate semiconductors and
equal to the Debye screening length \cite{Arg56}. Hence, in what
follows, the coefficient $k$ will be assumed to be independent of
the magnetic field $\mathbf{B}$.

Note that hybrid-impurity resonances can be observed only if all
hybrid-quantization levels are well-resolved and the photon
frequency $\omega$ is sufficiently monochromatic. We will assume
that the photon energy is considerably higher than the temperature
$T$ and that the collision width of the electron levels
$\hbar/\tau$ ($\tau$ is the relaxation time of the electron
momentum at scatterers) is small as compared to the temperature
and the photon energy $\hbar\omega$.

The impurity-unperturbed Hamiltonian of an electron in an
anisotropic parabolic quantum dot placed in an arbitrary directed
magnetic field $\mathbf{B}$ has the form
\begin{equation}
\label{ham} \hat
H=\frac{1}{2m^*}\left(\vec{p}-\frac{e}{c}\vec{A}\right)^2+\frac{m^*}{2}(\Omega^{2}_x
x^2 + \Omega^{2}_yy^2+\Omega^{2}_z z^2)\;,
\end{equation}
Here $\vec{A}=(B_yz/2-B_zy,0,B_xy-B_yx/2)$ is the vector potential
of the magnetic field $\vec{B}$.

A direct calculation of the matrix elements of electron-photon and
electron-impurity interaction is a complicate problem in our case
(it is not solvable analytically). However, the method of canonic
transformation of the phase space \cite{Gey05} allows us to
resolve this problem using only simple calculations from linear
algebra. In particular, in our preceding papers we used this
method to study hybrid and hybrid-phonon resonances in this system
\cite{Mar02,Gey01}. By means of a linear canonic transformation of
the phase space, we found the new phase coordinates
($\vec{P},\vec{Q}$) such that Hamiltonian (\ref{ham}) has the
canonic form \cite{Gey05,Gey01}
\begin{equation}
\label{hamnew} H=\frac{1}{2
m^*}(P^{2}_1+P^{2}_2+P^{2}_3)+\frac{m^*}{2}(\omega^{2}_1
Q^{2}_1+\omega^{2}_2Q^{2}_2+\omega^{2}_3Q^{2}_3)\;,
\end{equation}
where $\omega_i(i=1,2,3)$ are the hybrid frequencies depended on
the magnitude and direction magnetic field \cite{Gey05,Gey01}.

The spectrum of Hamiltonian (\ref{hamnew}) and, consequently, the
spectrum of Hamiltonian (\ref{ham}) has the form
\begin{equation}
\varepsilon_{nml}=\hbar\omega_1\left(n+\frac{1}{2}\right)+\hbar\omega_2\left(m+\frac{1}{2}\right)+\hbar\omega_3\left(l+\frac{1}{2}\right)\;,
\end{equation}
where $n,m,l=0,1,2\ldots$.

In our proceeding papers \cite{Gey05,Gey01} we have found the
transition matrix from the initial phase coordinates
($\vec{p},\vec{r}$) to the new ones ($\vec{P},\vec{Q}$). Using
this matrix we can easily calculate the matrix elements of the
coordinate and momentum operator because the wave function have a
simple form of the product of the oscillatory functions
$\Psi_{nml}=\Phi_n(Q_1)\Phi_m(Q_2)\Phi_l(Q_3)$ in the new phase
variables ($\vec{P},\vec{Q}$).
\section{Absorption coefficient}
Scattering by the impurity can lead to the two-stage transitions:
absorption of the photon is accompanied by the scattering from an
impurity or the scattering from an impurity is accompanied by the
absorption of the photon. The probability of these processes in
the second-order perturbation theory is determined by squared
$W_{\alpha\alpha'}$
\begin{eqnarray}
 \label{ver}
  W_{\alpha\alpha'}=\langle \alpha'|
\hat{H}_{eff}|\alpha\rangle=\sum_{\alpha''}\frac{\langle \alpha'|
\hat {H}_R|\alpha''\rangle\langle
\alpha''|\hat{V}|\alpha\rangle}{\varepsilon_{\alpha'}-\varepsilon_{\alpha''}
-\hbar\omega}+\sum_{\alpha''}\frac{\langle \alpha'| \hat
{V}|\alpha''\rangle\langle
\alpha''|\hat{H}_R|\alpha\rangle}{\varepsilon_{\alpha'}-\varepsilon_{\alpha''}+\hbar\omega}.
\nonumber\\
\end{eqnarray}
Here $\hat{H}_{eff}$ is the effective Hamiltonian of the
aforementioned processes, $\alpha=(n,m,l)$, $\alpha'=(n',m',l')$
are quantum numbers of initial and final states, $\hat{H}_R$ is
the operator of the electron-photon interaction, and $\hat{V}$ is
the operator of the electron-impurity interaction. In (\ref{ver}),
the first term describes processes involving, first, scattering
from an impurity and then, absorption of a photon; and the second
term accounts for the processes involving, first, absorption of a
photon and, then, scattering from an impurity.

Using the aforementioned method one can obtain the following
formula for the matrix elements of the operator of the
electron-photon interaction $\hat{H}_R$ \cite{Gey01}
\begin{eqnarray}
\langle nml|\hat{H}_R|n'm'l'\rangle=ie\hbar\sqrt{\frac{\pi
N_{\vec{f}}}{m^*\varepsilon\omega}} \left[X_1\sqrt{n+1}\delta_{n'
, n' -1 }\delta_{m , m'}\delta_{l,
l'}\right.\nonumber\\\left.+X_2\sqrt{m
+1}\delta_{n,n'}\delta_{m,m'-1}\delta_{l,l'}
+X_3\sqrt{l+1}\delta_{n,n'}\delta_{m,m'}\delta_{l,l'-1}\right].
\label{Xi}
\end{eqnarray}
Here $N_{\vec{f}}$ is the number of initial-state photons with
frequency $\omega$ and the coefficients $X_i$ ($i=1,2,3$) were
found in \cite{Gey01}. Note that relationship (\ref{Xi}) involves
only a term that corresponds to absorption of a photon.

Now we need to find the matrix elements of the electron-impurity
interaction operator. To calculate the matrix elements of
$\hat{V}$ it is conveniently to write the screened potential
$V(r)$ in the form of a Fourier series \cite{Zim60}
\begin{equation}
V(r)=\frac{4\pi
ze^2}{V_0\varepsilon}\sum_{\vec{q}}\frac{1}{q^2+k^2}e^{i\vec{q}\vec{r}}=
\sum_{\vec{q}}C_qe^{i\vec{q}\vec{r}}
\end{equation}
Here $V_0$ is the normalization volume and $C_q=4\pi
ze^2/V_0\varepsilon(q^2+k^2)$.

After simple, but rather cumbersome transformations, we obtain the
matrix elements for the operator $\hat{V}$
\begin{eqnarray}
\langle n'm'l'|\hat{V}|n''m''l''\rangle
&=&\sum_{\vec{q}}C_q\left(\frac{n''!m''!l''!}{n'!m'!l'!}\right)^{1/2}(-1)^{n'-n''}(-1)^{m'-m''}
(-1)^{l'-l''}\nonumber\\
&\times&
e^{-g^2/2}e^{-(\kappa_1\lambda_1+\kappa_2\lambda_2+\kappa_3\lambda_3)i/2}
e^{i\varphi_1(n'-n'')}e^{i\varphi_2(m'-m'')}e^{i\varphi_3(l'-l'')}\nonumber\\
&\times& L^{n'-n''}_{n''}(g^{2}_1)L^{m'-m''}_{m''}(g^{2}_2)
L^{l'-l''}_{l''}(g^{2}_3)g^{n'-n''}_{1}g^{m'-m''}_{2}g^{l'-l''}_{3}\,.
\label{V}
\end{eqnarray}
Here $l_i=\sqrt{\hbar/m^*\omega_i}$ ($i=1,2,3$) are the hybrid
length, $g_i=\sqrt{\lambda_i^2+\kappa_i^2l_i^4}/\sqrt{2}l_i$,
$\tan\varphi_i=\kappa_i l_i^2/\lambda_i$, $g^2=g_1^2+g_2^2+g_3^2$,
$L^{n'}_{n}(x)$ are the generalized Laguerre polynomials,
$\lambda_i=\hbar(b_{1i}q_x+b_{2i}q_y+b_{3i}q_z)$ ($i=1,2,3$),
$\kappa_{i-3}=b_{1i}q_x+b_{2i}q_y+b_{3i}q_z$ ($i=4,5,6$), $b_{ji}$
are components of the transition matrix from the phase variables
$(\vec{p},\vec{r})$ to $(\vec{P},\vec{Q})$ \cite{Mar02}.

Substituting the obtained expressions for the matrix elements of
$\hat{H}_R$ (\ref{Xi}) and $\hat{V}$ (\ref{V}) into (\ref{ver}),
 we get the following expression for the probability of transitions
\begin{eqnarray}
|W_{\alpha\alpha'}|^2&=&\frac{\pi e^2
N_{\vec{f}}}{m^*\varepsilon\omega}
\left(\frac{n'!m'!l'!}{n!m!l!}\right)\sum_{\vec{q}}e^{-g^2}g^{2(n'-n)}_1g^{2(m'-m)}_2g^{2(l'-l)}_3\nonumber\\
&\times&|C_q|^2|A(\omega)|^2
[L^{n'-n}_{n}(g^{2}_1)]^2[L^{m'-m}_{m}(g^{2}_2)]^2[L^{l'-l}_{l}(g^{2}_3)]^2\;,
\end{eqnarray}
where the factor
\begin{equation}
A(\omega)=\left(\frac{X_1g_1e^{i\varphi_1}}{\omega_1-\omega}+\frac{X_2g_2e^{i\varphi_2}}{\omega_2-\omega}+
\frac{X_3g_3e^{i\varphi_3}}{\omega_3-\omega}\right)\;.
\end{equation}
has the singularities in the points $\omega_i$.

 In the case of nondegenerate gas, the absorption coefficient
can be determined by the following formula \cite{Gey01,Bas65}
\begin{equation}
\Gamma(\omega)=\frac{2\pi\sqrt{\varepsilon}}{c\hbar
N_{\vec{f}}}\left(1-e^{-\hbar\omega/T}\right)\sum_{nml}\sum_{n'm'l'}f_0(\varepsilon_{nml})|W_{\alpha\alpha'}|^2\delta(\varepsilon_{nml}
-\varepsilon_{n'm'l'}+\hbar\omega)\;.
\end{equation}
Here $f_0(\varepsilon_{nml})$ is the electron distribution
function for the nondegenerate gas.

It is convenient to write the absorption coefficient as a sum of
partial absorption coefficients
\begin{equation}
\Gamma(\omega)=\sum_{nml}\sum_{n'm'l'}\Gamma(nml,n'm'l')\;,
\end{equation}
In our case
\begin{eqnarray}
\label{par}
\Gamma(nml,n'm'l')&=&\frac{2\pi^2e^2}{m^*c\hbar^2\sqrt{\varepsilon}\omega}\left(1-e^{-\hbar\omega/T}\right)
\left(\frac{n!m!l!}{n'!m'!l'!}\right)f_0(\varepsilon_{nml})\nonumber\\
&\times&\sum_{\vec{q}}e^{-g^2}g^{2(n'-n)}_1g^{2(m'-m)}_2g^{2(l'-l)}_3
|C_q|^2|A(\omega)|^2\nonumber\\
&\times&[L^{n'-n}_{n}(g^{2}_1)]^2[L^{m'-m}_{m}(g^{2}_2)]^2[L^{l'-l}_{l}(g^{2}_3)]^2\delta(\Delta\omega)\;,
\end{eqnarray}
where $
f_0(\varepsilon_{nml})=8n_0\sinh(\hbar\omega_1/2T)\sinh(\hbar\omega_2/2T)\sinh(\hbar\omega_3/2T)
\exp(-\varepsilon_{nml}/T)$, $N$ is the number of electron
density.

 In (\ref{par}) we introduced the
resonance detuning $
\Delta\omega=\omega_1(n-n')+\omega_2(m-m')+\omega_3(l-l')+\omega$.

Replacing the sum by the integral and substituting $C_q$ into
(\ref{par}), we can rewrite (\ref{par}) in the form
\begin{eqnarray}
\label{part} &&\Gamma(nml,n'm'l')=\frac{32\pi z^2e^6}{V_0
m^*c\hbar^2\varepsilon^{5/2}\omega}
f_0(\varepsilon_{nml})\left(1-e^{-\hbar\omega/T}\right)\delta(\Delta\omega)\nonumber\\
&\times&\left(\frac{n!m!l!}{n'!m'!l'!}\right)\int\limits_{\vec{q}}e^{-g^2}g^{2(n'-n)}_1g^{2(m'-m)}_2g^{2(l'-l)}_3
\frac{|A(\omega)|^2}{(q^2+k^2)^2}\nonumber\\
&\times&
[L^{n'-n}_{n}(y^{2}_1)]^2[L^{m'-m}_{m}(y^{2}_2)]^2[L^{l'-l}_{l}(y^{2}_3)]^2d^3q
\;.
\end{eqnarray}

 Equation (\ref{part}) clearly shows that the partial
coefficients $\Gamma(nml,n'm'l')$ have delta-function
singularities at the points where $\delta(\Delta\omega)=0$.

Transitions from the ground state will provide the major
contribution to the absorption coefficient. Then we can write the
partial coefficients in the following form (taking into account
that for a degenerate gas at low temperatures, the distribution
function may be assumed to be $f_0(\varepsilon_{nml})\approx1$ and
$\hbar\omega\gg T$)
\begin{eqnarray}
\label{ground} &&\Gamma(000,n'm'l')=\frac{32\pi z^2e^6n_0}{
V_0m^*c\hbar^2\varepsilon^{5/2}\omega}
\frac{\delta(\Delta\omega)}{n'!m'!l'!}\int\limits_{\vec{q}}e^{-g^2}g^{2n'}_1g^{2m'}_2g^{2l'}_3
\frac{|A(\omega)|^2}{(q^2+k^2)^2}d^3q
\;.\nonumber\\
\end{eqnarray}
\section{Results and discussions}
For the qualitative analysis of the absorption it is convenient,
for definiteness, assume that the magnetic field is directed along
$y$-axis. Then it is easy to show that the transitions between the
levels with the different $n$ and $l$ are forbidden. In this case
only the coefficient $b_{22}$ is different from zero
\begin{eqnarray}
b_{22}=\frac{\left( \Omega_x^2-\omega_2^2\right)\left(
\Omega_z^2-\omega_2^2\right)-\omega_c^2\omega_2^2}{m^*\omega_2
\left\{\omega_c^4\omega_2^4+\left[\left(\Omega_x^2-\omega_2^2\right)\left(
\Omega_z^2-\omega_2^2\right)-\omega_c^2\omega_2^2\right]^2\right\}^{1/2}}\;
(i=1,2,3)\;,
\end{eqnarray}
where the characteristic frequencies is determined by the
following formulas
\begin{eqnarray}
\omega_{1,3}=\frac{1}{2}\left[\sqrt{\left(\Omega_x+\Omega_z
\right)^2+\omega_c^2}\pm\sqrt{\left(\Omega_x-\Omega_z
\right)^2+\omega_c^2} \right], \omega_2=\Omega_y\;.
\end{eqnarray}
Here $\omega_c=eB/m^*c$ is the cyclotron frequency.

Note that hybrid-impurity resonances arise in the points
$\omega=\omega_2 (n'-n)\equiv\Omega_y(n'-n)$. Hence the resonance
frequency is independent of the magnetic field in this case.
\begin{figure}
\begin{center}
\includegraphics*[width=10cm]{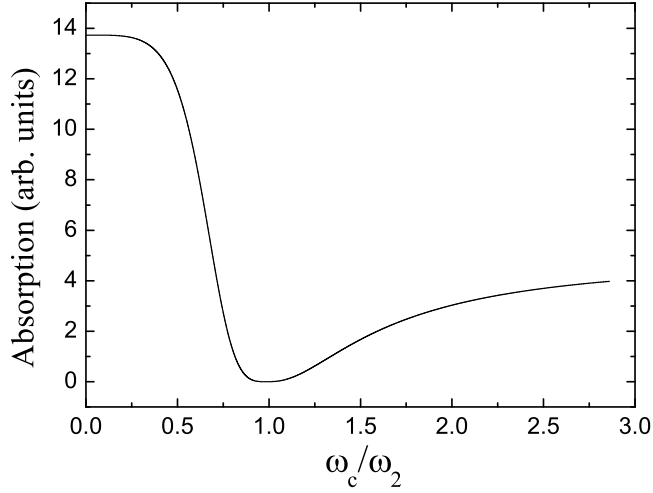}
\end{center}
\caption{The absorption coefficient as a function of the magnetic
field in the case of transitions
$|0,0,0\rangle\rightarrow|0,2,0\rangle$.
$\omega=1.0\times10^{13}$s$^{-1}$,
$\Omega_x=1.2\times10^{12}$s$^{-1}$,
$\Omega_y=4.1\times10^{13}$s$^{-1}$,
$\Omega_z=7.3\times10^{12}$s$^{-1}$.} \label{fig:1}
\end{figure}
\begin{figure}
\begin{center}
\includegraphics*[width=10cm]{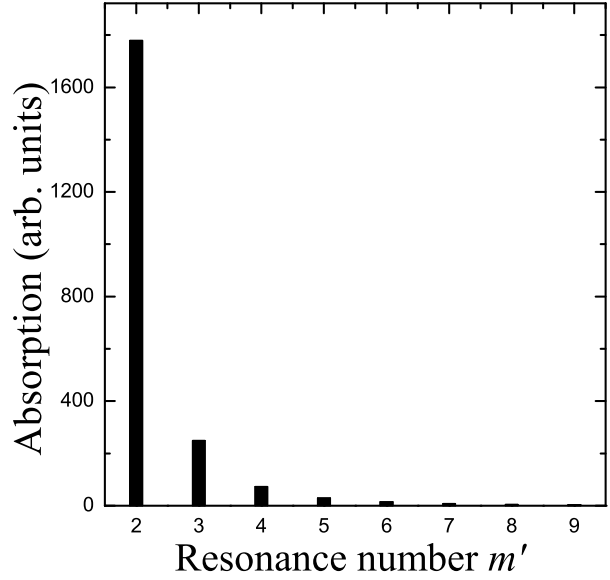}
\end{center}
\caption{The relative intensity of the absorption peaks at the
resonance points $\omega=m'\omega_2$.
$\Omega_x=2.4\times10^{12}$s$^{-1}$,
$\Omega_y=3.0\times10^{12}$s$^{-1}$,
$\Omega_z=7.3\times10^{13}$s$^{-1}$.} \label{fig:2}
\end{figure}
If we know $b_{ji}$ it is easy to obtain the formulas for
$A(\omega)$, $g^2$ and $X_2$ (we assume that the polarization
vector is parallel to the magnetic field)
\begin{eqnarray}
A(\omega)=\frac{X_2}{\omega-\omega_2}g_0q_y\;, g^2=g_0^2q_y^2\;,
X_{2}^2=\frac{2\omega_2g_0^2}{l_2^2}\;, X_1=X_3=0\;,
\end{eqnarray}
where $g_0=\hbar b_{22}/\sqrt{2}l_2$. Substituting these formulas
into (\ref{ground}), we get
\begin{eqnarray}
&&\Gamma(000,0m'0)=\frac{32\pi z^2e^6n_0}{
V_0m^*c\hbar^2\varepsilon^{5/2}\omega}\frac{g_0^{2(m'+1)}X_2^2}{(\omega-\omega_2)^2}
\frac{1}{!m'!}\int\limits_{\vec{q}}e^{-g_0^2q_y^2}
\frac{q_y^{2(m'+1)}}{(q^2+k^2)^2}d^3q
\delta(\Delta\omega)\;.\nonumber\\
\end{eqnarray}
Integrals over $q_x$ and $q_z$ can be easily evaluate. As a result
we get
\begin{eqnarray}
\label{end} &&\Gamma(000,0m'0)=\frac{128\pi^2z^2e^6n_0\omega_2}{
V_0m^*c\hbar^2\varepsilon^{5/2}l_2\omega}\frac{|g_0^3|}{(\omega-\omega_2)^2}
\frac{1}{!m'!}\int\limits_{0}^{\infty}
\frac{e^{-x^2}x^{2(m'+1)}}{x^2+g_0^2k^2}dx
\delta(\Delta\omega)\;.\nonumber\\
\end{eqnarray}
The analysis of Eq.(\ref{end}) shows that the partial absorption
coefficients as a function of a magnetic field have minimum in the
points where the magnetic field satisfies condition
$\omega_c=\Omega_y$ (Fig.\ref{fig:1}). It is important that
changing the magnitude of the magnetic field we can strongly
decrease losses of electromagnetic radiation in the quantum dots.
In its turn the intensity of the resonance peaks falls off rapidly
with increasing of the number of resonance level $m'$. In
particular, as one can see from Fig.\ref{fig:2} the absorption
peak with $m'=2$ by an order of magnitude greater than one with
$m'=3$.

In the general case the partial absorption coefficients
(\ref{part}) have singularities of two types. The first type of
resonances is stipulated by the frequency factor $A(\omega)$ in
the absorption coefficient. In this case singularities arise at
the points $\omega_i$ ($i=1,2,3$). It corresponds to conventional
resonances in intraband absorption in the absence of scattering.
This is an analog of a hybrid resonance in this system which was
studied by the authors in Ref.\cite{Gey01}. Singularities of the
second type correspond to the hybrid-impurity resonance. In this
case the resonance peaks arise at harmonics of the hybrid
resonances in the points
$\omega=\omega_1(n'-n)+\omega_2(m'-m)+\omega_3(l'-l)$ due to the
selection rules for the transitions in second-order perturbation
theory and the law of energy conservation in such transitions. The
position of the hybrid-impurity resonances depends strongly on the
magnetic field and the characteristic frequencies of the parabolic
confinement. It follows from the corresponding dependence of the
hybrid frequencies. We stress the point that the scattering by
impurities removes the forbiddennes from the transitions between
levels different from the neighboring ones even if if we have only
one impurity in the quantum dot. Therefore, the resonances arising
at harmonics of conventional hybrid resonance can be identified as
a hybrid-impurity resonance at ionized impurity.

\ack The present work was supported by the Russian Foundation for
Basic Research (05-02-16145) and the Grant of President of Russia
for Young Scientists (MK-2062.2008.2).

\end{document}